\documentclass[twocolumn,showpacs,amsmath,amssymb,floatfix,prl,aps,superscriptaddress,showkeys]{revtex4}

\usepackage{graphicx}% Include figure files
\usepackage{epsfig}
\usepackage{dcolumn}% Align table columns on decimal point
\usepackage{bm}% bold math
\usepackage{times}
\usepackage{color}
\usepackage{soul}

\begin{document}
\title{The random walk of an electrostatic field using parallel infinite charged planes}
\author{Rodrigo Aldana}
\address{Instituto Tecnol\'ogico y de Estudios Superiores de Occidente, Perif\'erico Sur Manuel G\'omez Mor{\'i}n 8585 C.P. 45604, Tlaquepaque, Jal., MEXICO}
\author{Jos\'e Vidal Alcal\'a}
\address{CIMAT y C\'atedras CONACYT, Perif\'erico Norte 13615, 97119 M\'erida, Yucat\'an, MEXICO}
\author{Gabriel Gonz\'alez}\email{gabriel.gonzalez@uaslp.mx}
\affiliation{C\'atedras CONACYT, Universidad Aut\'onoma de San Luis Potos\'i, San Luis Potos\'i, 78000 MEXICO}
\affiliation{Coordinaci\'on para la Innovaci\'on y la Aplicaci\'on de la Ciencia y la Tecnolog\'ia, Universidad Aut\'onoma de San Luis Potos\'i,San Luis Potos\'i, 78000 MEXICO}
\begin{abstract}
We show that it is possible to generate a random walk with an electrostatic field by means of several parallel infinite charged planes in which the surface charge distribution could be either $\pm\sigma$. We formulate the problem of this stochastic process by using a rate equation for the most probable value for the electrostatic field subject to the appropriate transition probabilities according to the electrostatic boundary conditions. Our model gives rise to a stochastic law when the charge distribution is not deterministic. The probability distribution of the electrostatic field intensity, the mean value of the electrostatic force and the energy density are obtained. 
\end{abstract}

\maketitle

\section{I. Introduction}
\label{sec:introduction}
In a typical electrostatic problem you are given a volume charge distribution $\rho({\bf r})$ and you want to find the electric field ${\bf E}$ it produces. The fundamental relation between the source of the field, i.e. charge distribution, and the electric field is given by Gauss's law \cite{Griffiths}
\begin{equation}
\nabla\cdot{\bf E}=\frac{\rho}{\epsilon_0}
\label{eq001}
\end{equation}
where $\epsilon_0$ is the electric permittivity of free space. For the particular case of infinite charge sheets the electric field depends on one variable only, i.e. ${\bf E}(z)$, and the field strength it produces is constant for all distances from the plane. The electric field caused by an infinite sheet of volume charge density $\rho(z)=\sigma\delta(z)$ is given by \cite{Sadiku}
\begin{equation}
{\bf E}=\frac{\sigma}{2\epsilon_0}\mbox{sign}(z) 
\label{eq002}
\end{equation}
where $\sigma$ is the surface charge density and the function sign is defined as \cite{Weber}
\begin{equation}
\mbox{sign}(z)=\left\{ \begin{array}{ll}
+1 & \mbox{if $z>0$}\\
-1 & \mbox{if $z<0$},
\end{array}\right.
\label{eq003}
\end{equation}
and has been introduced to account for the vector nature of the electric field. The total electric field due to $N$ of these infinite charge planes is then just the sum of the individual contributions, i.e.
\begin{equation}
{\bf E}=\sum_{n=1}^{N}\frac{\sigma(z_n)}{2\epsilon_0}\mbox{sign}(z-z_n),
\label{eq004}
\end{equation}
where $\sigma(z_n)$ represents the surface charge density of the $n$-th charge plane at position $z_n$. Note that it is crucial to know the surface charge density on each infinite sheet to completely determine the value of the electric field through Eq. (\ref{eq004}), if the source of the electric field is not known then we would need to approach the problem in a different way. We can formulate this same problem from a probabilistic point of view using random walks. The connection between electrostatics and Brownian motion or random walk was first shown in 1944 by Kakutani \cite{kakut}. Since then many investigators have applied the probabilistic potential-theory to solve differential and integral equations for calculating electrostatic field and capacitance in integrated circuits \cite{yu,yu1}. More recently, the connection between electrostatics and quantum mechanics was first shown by one of the authors \cite{gg,gg1}.
The purpose of this paper is to give a description of the electrostatic field when the source of the field is not known. If the volume charge density is not known then we can not use Eq. (\ref{eq001}) to solve for the electric field, instead we need to use the boundary conditions for the electric field. For the restricted case of electrostatic charges an fields in vacuum, the appropriate boundary conditions for the electric field across a surface charge distribution is given by \cite{Griffiths}
\begin{equation}
			{\bf E}_{above}-{\bf E}_{below}=\frac{\sigma}{\epsilon_0}\hat{{\bf n}}
			\label{EqBoundary}
\end{equation}
where $\hat{{\bf n}}$ is a unit vector perpendicular to the surface. For the case of parallel infinite charged planes the boundary condition is given by
\begin{equation}
			E(z)_{above}-E(z)_{below}=\sigma/\epsilon_0
			\label{eq02}
\end{equation}
Equation (\ref{eq02}) says that, for the special case in which the infinite charged planes have a surface charge distribution given by $\sigma$, the future electric field value depends only on the present electric field value. This means that given precise information of the present state of the electric field value, the future electric field value does not depend on the past history of the process. This point is at the very heart of the Markov chain processes. This means that we can use a probabilistic model which gives a complete description of the electric field when the source of the field is not known.  \\
The problem that we would like to address in this paper is the one where you have several parallel infinite charged planes placed along the $z$ axis in which the surface charge density could be either $\pm\sigma$. Then the field would evolve along the $z$ axis making random jumps each time it crosses an infinite charge sheet.  We are going to show that the behavior of the electric field can be studied like a random walk. The motivation of this paper is due to the fact that many physical systems display a non-deterministic disorder. A better understanding and prediction of the nature of these systems is achieved by considering the medium to be random. The randomness models the effect of impurities on a physical system or fluctuations. Randomness appears at two levels in our problem. It comes in the description of the intensity of the electric field and also comes in the description of the medium.

The article is organized as follows. In section II we give a brief review of Markov chains. Then in section III we will apply the Markov chain theory to the case where we have several infinite charged planes placed along the $z$ axis in which  the surface charge density can be either $\pm \sigma$. In section IV we present the numerical results and use our model to calculate the mean electrostatic force and energy density for our problem. In the last section we summarize our conclusions.

%%%%%%%%%%%%%%%%%%%%%%%%%%%%%%%%%%%%%%%%%%%%%%%%%%%%%%%%%%%%%%%%%%%%%%%%%%%%%%
%Agregar referencia 
\section{II. Markov chains}
\label{sec:RandomWalk}
In this section we will give a brief explanation of the Markov chain theory concepts that we need for the formulation of the problem. To introduce these concepts lets imagine a random experiment like throwing a normal dice. In this case we have six possible outcomes which are $1,2,3,4,5$ or $6$. We will denote the set of the possible outcomes with the letter $S$ and name it as the {\it state space}. These outcomes define a random variable $X$ which can take values from $S$.
Each outcome can be associated with their respective probability $p$ which must take values only in the interval $0<p<1$. We can write $P(X = i) = p_i$ which means that the probability of the random variable $X$ to take a value $i$ its equal to $p_i$ with $i \in S$ \cite{Leong}.  

Also if the same experiment is repeated $n$ times, and in general each trial is dependent of the previous trials then $p_i(n)$ represents the state of the random variable $X_n$ after $n$ trials. If the state at a trial $n$ is $i$, the probability that the next state is equal to $j$  is defined by \cite{Rincon,DTGillespie}

\begin{equation}
\begin{split}
			P(X_{n+1}=j|X_n=i_n,X_{n-1}=i_{n-1},\dots,X_{0} = i_0)\\ 
			= P(X_{n+1} = j | X_{n} = i) = p_{ij}
			\label{eqMarkovProp}
\end{split}
\end{equation}

Equation (\ref{eqMarkovProp}) is known as the Markov property and says that the state of a random variable after $n+1$ transitions, only depends on the state of the random variable after $n$ transitions. In other words, the future of the system will only depend on the present state \cite{Bertsekas,DTGillespie}. 

If we want to know the probabilities of a single state it is useful to use the total probability theorem which is given by \cite{Bertsekas} 
\begin{equation}
			P(B) = \sum_{i} P(B|A_i)P(A_i)
			\label{totalprob}
\end{equation}

Where the index $i$ runs for all the states $A_i$ that $B$ depends on. If $B$ is the event of $X_{n+1} = j$ then $P(X_{n+1} = j) = p_j(n+1)$. $A_i$ is the event of $X_{n} = i$, thus $P(X_{n} = i) = p_i(n)$, then the theorem transforms into the following equation \cite{Wilson,Karlin}

\begin{equation}
			p_j(n+1) = \sum_{i} p_{ij}(n)p_i(n)
			\label{markovprod}
\end{equation}

This notation allows us to interpret equation (\ref{markovprod}) as a multiplication of a horizontal vector formed with the probabilities of some certain state with a matrix formed by the $p_{ij}$. So if there are $N$ possible states, $\mathbf{p}(n) = [p_{i_0}(n),\dots,p_{i_{N-1}}(n)]$ and $\mathbf{W}$ is the matrix with elements $p_{ij}$, the relation between states can be written as \cite{Rincon,Wilson}
\begin{equation}
			\mathbf{p}(n+1) =\mathbf{p}(n)\mathbf{W}(n)
			\label{matMarkov}
\end{equation}
 
Given the initial condition $\mathbf{p}(0)$ we can determine the solution to Eq. (\ref{matMarkov}) \cite{Kemeny}.
%%%%%%%%%%%%%%%%%%%%%%%%%%%%%%%%%%%%%%%%%%%%%%%%%%%%%%%%%%%%%%%%%%%%%%%%%%%%%%

\section{III. The Model}
\label{sec:model}
Suppose that we are given a collection of $N$ infinite charged planes, half of them have a constant charge distribution $\sigma$ and the other half have a constant charge distribution $-\sigma$. We can write this condition as

\begin{equation}
	\sum_{n=0}^{N-1} \sigma (z_{n}) = 0
	\label{neutralcond}
\end{equation}

If we randomly place the $N$ infinite charged planes parallel to each other along the $z$ axis at position $z_n = n\in \{0,1,\dots,N-1\}$, there is no way to know the values of the electric field in between the planes unless we measure the charge on the planes. Instead of that we will use Markov chain theory to analyze the problem.

\begin{figure}[hb]
			\centering
			\includegraphics[width=0.3\textwidth]{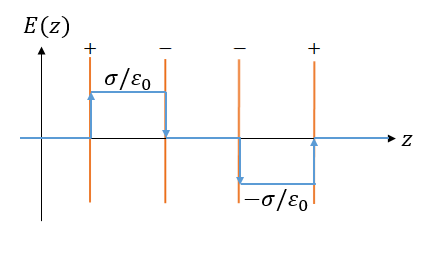}
			\caption{Values for E(z) inside a given configuration of charged planes} 
			\label{fig:field2}
	\end{figure}

We will first consider a system of four charged planes. $E(z)$ equals zero at the left and right side of the configuration due to the neutrality of  the system \cite{Griffiths}. After crossing the first charged plane, $E(z)$ would increase or decrease in an amount of $\sigma/\epsilon_0$ depending if the first charged plane had a positive or negative charge, and the same for the remaining planes according to equation (\ref{eq02}). In figure \ref{fig:field2} there is an example of how the electric field behaves for a known charge density. 
		\begin{figure}[hb]
			\centering
			\includegraphics[width=0.3\textwidth]{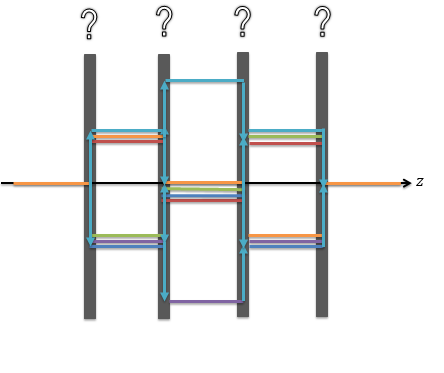}
			\caption{Possible values for E(z) inside a four charged planes system where the charge distribution is not known}
			\label{fig:field1}
	\end{figure}
In figure \ref{fig:field1} we show the possible values the electric field can take when the charge density is not known.\\ 
	
For the sake of simplicity, we normalized the electric field dividing  by $\sigma/\epsilon_0$ so
\begin{equation}
 \bar{E}(z) = \frac{\epsilon_0}{\sigma}E(z)
\label{NormalE}
\end{equation}
Due to this normalization $\bar{E}(z)$ takes only integer values $i$. Defining $\bar{E}_i(n)$ as our random variable, where $n$ is the region between $z_n$ and $z_{n+1}$, our state space $S$ is defined by integer values between the interval $[-N/2,N/2]$. 
To calculate the transition probabilities $p_{ij}$ we use the fact that the boundary condition in equation (\ref{eq02}) prevent us to have a value bigger than $i+1$ or smaller than $i-1$ at a place $z = n+1$ when $\bar{E}_i(n) = i$. This results in $p_{ij} = 0$ for $j>i+1$ and for $j<i+1$. After crossing a charged plane the field has to have a different value, so $p_{ii} = 0$. The last values for $p_{ij}$ are determined by the probability of finding a negative or positive charged plane at the transition point. Defining $W_{\pm}(n)$ as the probability to find a positive or negative charged plane at a place $z_n = n$, we get the following results for $p_{ij}$

\begin{equation}
		p_{ij}=\left\{ 
		\begin{array}{ll}
			W_{+}(n) & \mbox{if $j=i+1$}\\
			W_{-}(n) & \mbox{if $j=i-1$}\\
			0        & \mbox{otherwise}
		\end{array}\right.
	\label{pij}
\end{equation}

A simple way to calculate the factors $W_{\pm}(n)$ its by dividing the remaining positive or negative planes by the remaining total planes. Let $p_{\pm}$ be the number of positive (negative) planes between $0 \leq z \leq n$, then $p_+ + p_- = n$.  Also the value of the electric field $i$ at a point $n$ is completely defined by the amount of positive and negative plates that have left behind, therefore $p_+ - p_- = i$. Using these two relations we can solve for $p_+$:
	\begin{equation}
		p_{+} = \frac{n+i}{2}
		\label{eq06}
	\end{equation}
The remaining positive planes are $N/2 - p_+$ and the remaining total planes are $N-n$, then $W_{+}$ is given by
	\begin{equation}
		W_{+}(n)  =  \frac{\frac{N}{2}-\frac{i+n}{2}}{N-n}=\frac{1}{2}(1-\frac{i}{N-n})
	\label{eqWp}
	\end{equation}
and $W_{-}(n) = 1 - W_{+}(n)$.

We known that the system is neutral, so the electric field must be zero at $z = 0$. In other words the probability for $\bar{E}$ to be zero at $z=0$ equals one, and the other cases have $0$ probability. Using our notation the last statement can be written as $p_0(0)=1$ and $p_i(0) = 0$ for $i \neq j$.  

With all these information we can use Eq. (\ref{matMarkov}) to obtain the most probable value of the electrostatic field in between the charged planes.
	
%%%%%%%%%%%%%%%%%%%%%%%%%%%%%%%%%%%%%%%%%%%%%%%%%%%%%%%%%%%%%%%%%%%%%%%%%%%%%%%
\section{IV. Results}
\begin{figure}
			\centering
			\includegraphics[width=0.4\textwidth]{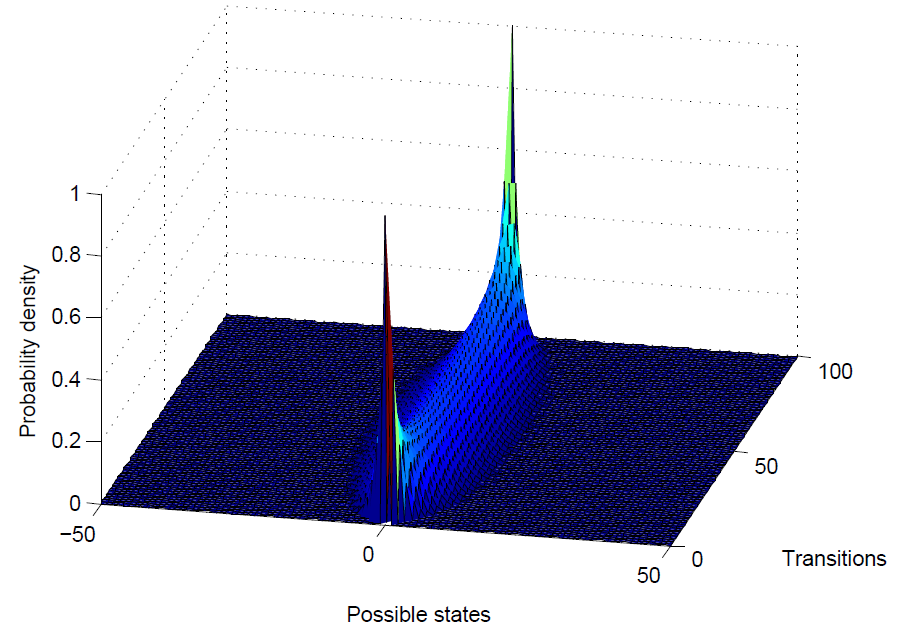}
			\caption{State probabilities for every transition}
			\label{fig:PMFcomplete}
	\end{figure}

In this section we present our results for a the system which consists of $100$ charged planes ($N=100$).Figure \ref{fig:PMFcomplete} shows a 3D plot for $\mathbf{p}(n)$. Note that the highest values for the probability are reached at $n = 0$ and $n = N-1$ with $i = 0$, which shows the fact that the electric field must be zero at these points. In the direction of the increasing $n$ the probabilities spreads out on the possible states, so the electric field becomes more uncertain at places near the middle of the configuration. After that point, the probabilities change to finally reach $1$ at $n = N-1$.

\begin{figure}
			\centering
			\includegraphics[width=0.4\textwidth]{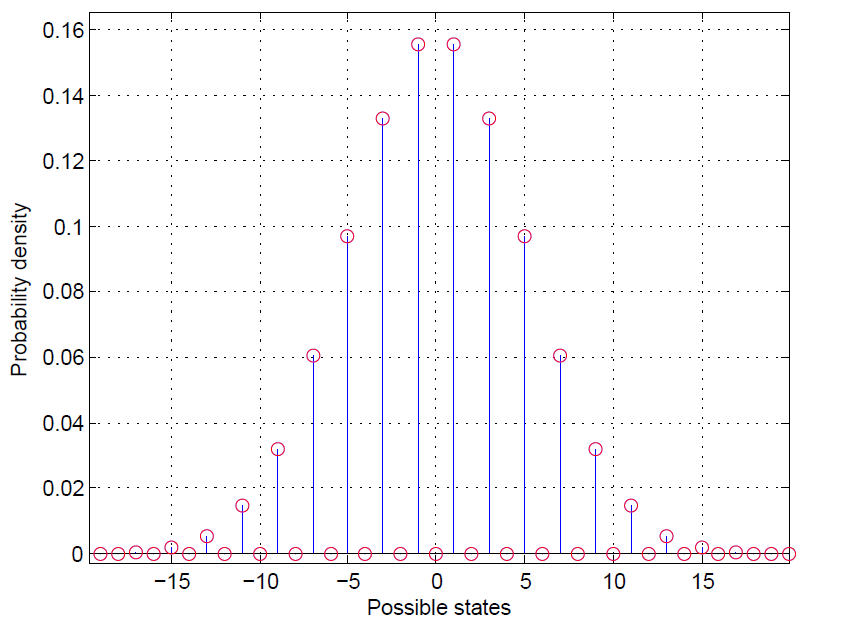}
			\caption{State probabilities after 50 transitions}
			\label{fig:PMFsection}
	\end{figure}

Its important to note that these probability functions for each transition are discontinuous functions. The plot in figure \ref{fig:PMFsection} represents a cross section for $\mathbf{p}(n)$ at the middle of the configuration ($n = 50$). In this plot it is easy to see that there are jumps between probability zero and nonzero probability along the possible states. This is due to the boundary condition given in Eq. (\ref{EqBoundary}). Note that the field cannot remain the same so there are forbidden states at a certain transition $n$.\\

Now we can calculate the two physical quantities that we are interested in which are the mean electrostatic force and mean electrostatic energy density. It is well known that the electrostatic force is given by $\bm{F} = q \bm{E}$ and the energy density by $\delta_E = \epsilon_0 E^{2}(z)/2$. In order to calculate these two quantities we need to calculate the first and second moments for the electrostatic field, i.e. $\langle \bar{E}(n) \rangle$ and  $\langle \bar{E}^2(n) \rangle$. To calculate such moments we use the {\it k}-th moment formula \cite{Karlin, Bertsekas}

	\begin{equation}
		\langle \bar{E}^k(n) \rangle = \sum_{i} i^k p_i(n)
		\label{kthmoment}
	\end{equation}
	
\begin{figure}
			\centering
			\includegraphics[width=0.2\textwidth]{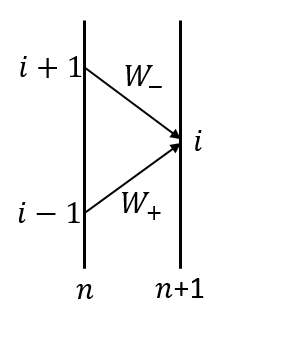}
			\caption{Relation between transition $n$ and $n+1$}
			\label{fig:moments}
	\end{figure}

Where the index $i$ runs through all $S$. In order to calculate these moments let us rewrite the form of equation (\ref{matMarkov}) as follows. Most of the probabilities $p_{ij}$ equals to zero, so the probability $p_i(n+1)$ can be calculated adding the 2 possible paths from the last transition given by the possible values at the transition $n$. Each of the paths may me multiplied by the factors $W_{\pm}(n)$ \cite{Gillespie} as is depicted in figure (\ref{fig:moments}). 
So the new form of equation (\ref{matMarkov}) is given by
\begin{equation}
		\begin{array}{lll}
		p_i(n+1)  &= &p_{i+1}(n)W_-(i+1,n) \\
		         &+ &p_{i-1}(n)W_+(i-1,n) 
		\end{array}
		\label{moment1}
\end{equation}
Multiplying both sides of equation (\ref{moment1}) by $i^k$ and summing from $-N/2$ to $N/2$ we may use the definition given in (\ref{kthmoment}) to obtain 
	\begin{equation}
		\begin{array}{ll}
		\langle \bar{E}^k(n+1) \rangle &= \displaystyle\sum_{i=-N/2}^{N/2} i^{k}p_{i+1}(n)W_-(i+1,n)\\
													 &+ \displaystyle\sum_{i=-N/2}^{N/2}i^{k}p_{i-1}(n)W_+(i-1,n) 
				\end{array}
		\label{moment2}
	\end{equation}
Its important to note that $p_{N/2+1}(n) = p_{-N/2-1}(n) = 0$ because $N/2+1$ and $-N/2-1$ does not exist in $S$. Also $W_{\pm}(\pm N/2,n) = 0$ because the field cannot get a value higher (lower) than $\pm N/2$. Using these facts and setting $k=1$ in (\ref{moment2}) we obtain
	\begin{equation}
		\langle \bar{E}(n+1) \rangle = \frac{N-n-1}{N-n}\langle \bar{E}(n) \rangle
		\label{diference1}
	\end{equation}
Due to the neutrality of the system we can write $\langle \bar{E}(0) \rangle = 0$. Using this initial value and substituting it in (\ref{diference1}) its obvious that $\langle \bar{E}(n) \rangle = 0 $ for all $n$ inside the plane configuration as expected \cite{Grimaldi}. Therefore the mean force is zero everywhere. 

Setting $k=2$ in (\ref{moment2}) and doing a little bit of algebra, equation (\ref{moment2}) turns into
	\begin{equation}
		\langle \bar{E}^2(n+1) \rangle = \frac{N-n-2}{N-n}\langle \bar{E}^2(n) \rangle +1
		\label{diference2}
	\end{equation}
Let $A(n)$ be the factor that multiplies $\langle \bar{E}^2(n) \rangle$ in equation (\ref{diference1}), and knowing that $\langle \bar{E}^2(0) \rangle = 0$ and using recursively the difference equation in (\ref{diference2}) \cite{Grimaldi}, the states for the second moment are given by
	\begin{equation}
		\begin{array}{l}
		\langle \bar{E}^2(1) \rangle = 1\\
		\langle \bar{E}^2(2) \rangle = A(1)+1\\
		\langle \bar{E}^2(3) \rangle = A(2)A(1)+A(2)+1\\
		\dots \\
		\langle \bar{E}^2(n) \rangle = 1+\displaystyle\sum_{i=1}^{n-1}\prod_{j=1}^{i}A(n-j)
		\end{array}
		\label{2moment}
	\end{equation}
Substituting $A(n)$ in equation (\ref{2moment}), the second moment is given by
	\begin{equation}
	\langle \bar{E}^2(n) \rangle = 1+\displaystyle\sum_{i=1}^{n-1}\frac{(N-n)(N-n-1)}{(N-n+i)(N-n+i-1)}
		\label{final2moment}
	\end{equation}
	
\begin{figure}
			\centering
			\includegraphics[width=0.4\textwidth]{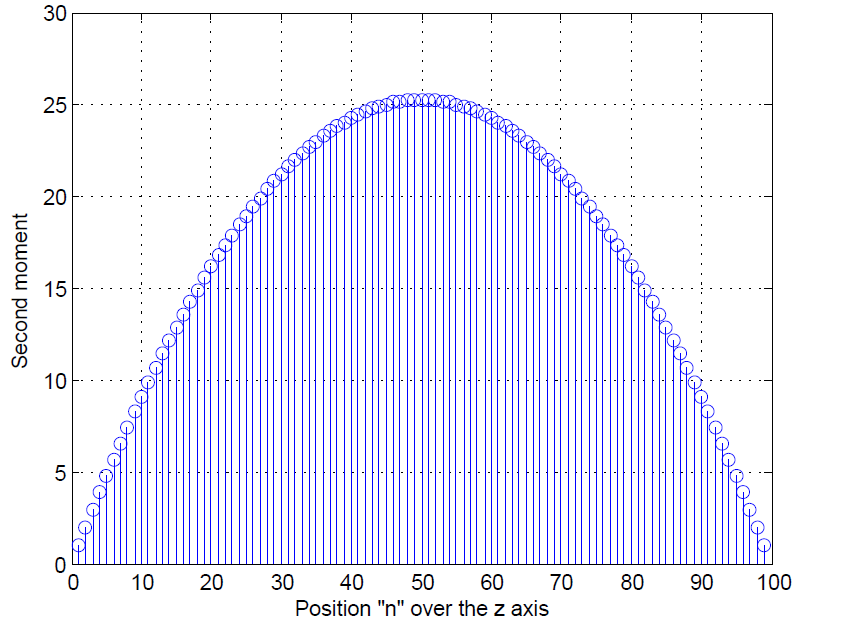}
			\caption{Evolution of the second moment along the $z$ axis}
			\label{fig:momentfig}
	\end{figure}

	To see how the energy evolves along the $z$ axis we used the same example with $N = 100$ as before and numerically calculated the energy using (\ref{final2moment}).The result is depicted in figure (\ref{fig:momentfig}). 
The the mean energy may be calculated using the definition of our random variable given in (\ref{NormalE})  as 
	\begin{equation}
		\langle \delta_E \rangle = \frac{\epsilon_0}{2}\langle E^2(n)\rangle = \frac{\sigma^2}{2\epsilon_0}\langle \bar{E}^2(n)\rangle
		\label{Energy}
	\end{equation}
We can use standard probability theory to validate our calculations for the example given above. For the case of $N=100$ there is a probability of $50!50!/100!$ that the electrostatic field reaches the maximum (minimum) value $\pm 50$ after $50$ transitions. Using our approach this means that the probability to get a value of $50$ is obtained by
\begin{equation}
p_{50}(50) = \prod_{j = 1}^{50} {W_{+}(50-j,50-j)}
\label{eqval1}
\end{equation}  
Equation (\ref{eqval1}) means that we need $50$ positive charged planes in order to reach the maximum value of the electrostatic field. Using the fact that $W_+(i,n)=(1-i/(N-n))/2$ we get $W_+(50-j,50-j)=j/(50+j)$, inserting this into Eq.(\ref{eqval1}) we get
\begin{equation}
p_{50}(50) = \prod_{j = 1}^{50}\frac{j}{50+j}=\frac{(1)(2)(3)\cdots(49)(50)}{(51)(52)(53)\cdots(99)(100)} = \frac{50!50!}{(100)!}
\label{eqval2}
\end{equation} 
Equation (\ref{eqval2}) shows that our result is consistent with the standard probability theory.
\section{V. Conclusions} 
We have shown that it is possible to perform a random walk with and electrostatic field by means of several parallel infinite charged planes in which the surface charge distribution is not explicitly known. We have worked out the special case where the charged planes have a constant surface charge distribution $\pm\sigma$ and the overall electrostatic system is neutral, i.e. there is the same number of positive and negative charged planes. We use Markov chain theory in order to give the most probable value for the electrostatic field in between the charged planes and use these results to obtain the mean electrostatic force and energy density. Our model gives rise to a stochastic law when the charge distribution is not deterministic.  
\section{Acknowledgments}
This work was supported by the program ``C\'atedras CONACYT".

\end{document}